*Szabolcs Nagy – Gergő Hajdú*

# The relationship between content marketing and the traditional marketing communication tools

*Digitalization is making a significant impact on marketing. New marketing approaches and tools are emerging which are not always clearly categorised. This article seeks to investigate the relationship between one of the novel marketing tools, content marketing, and the five elements of the traditional marketing communication mix. Based on an extensive literature review, this paper analyses the main differences and similarities between them. This article aims to generate a debate on the status of content marketing. According to the authors' opinion, content marketing can be considered as the sixth marketing communication mix element. However, further research is needed to fill in the existing knowledge gap.*
*Keywords: content marketing, trends, advertising, sales promotion, direct marketing, personal selling, public relations*
*JEL: M31, M37*

**Introduction**

Digitalization and the ongoing information technology revolution pose remarkable possibilities and challenges for marketing (Piskóti, 2018). Due to digitalization, consumer behaviour is constantly changing. Consumers' stimulus threshold is increasing because of the greater exposure to information (Kotler et al., 2017). At the same time, smart devices are becoming increasingly dominant (Nagy, 2017). E-commerce (Nagy, 2016), and the various social networks are becoming popular (Sethi, 2018). These trends accelerate the emergence of new methods and trends in marketing (Nagy, 2020). It is advisable for marketers to understand how those new methods and tools work since they help to reach out to consumers to influence their behaviour. However, the lack of advanced information technology in Hungary poses some problems in this process (Kamaraonline, 2018).
Marketing communication tools can often be divided into two main groups. Traditional and digital solutions can be distinguished. However, according to Kotler et al. (2017), the two categories have recently been merging. Content marketing is essentially a digital solution having some offline features as well. The significance of content marketing is supported by Kotler et al. (2017), who found - referring to the research findings of Content Marketing Institute and the MarketingProfs - that 76% of the B2C companies and 88% of the B2B companies used content marketing in North America. Furthermore, B2C companies spent 32% of their marketing budgets on content marketing, while B2B companies spent 28%. 57% of the B2C companies increased their content marketing budget by at least 1%, while 29% of them did not change the budget (Brenner 2019, based on Content Marketing Institute 2019). The companies mainly increased their content marketing budgets in the following areas: content production (56%), content marketing personnel (37%), paid distribution of content (36%), content marketing technology (29%), and content marketing outsourcing (29%) (Murton Beets 2018). These facts also underline the importance of content marketing in today's digital world.
If we accept that traditional and digital solutions have been merging (Kotler et al., 2017), it means that the traditional classification of marketing communication tools should be revised. The communication tools should rather be classified according to their functions and operating mechanisms than according to the type of technological solutions. From this perspective, *content marketing (CM) is a new approach to marketing communication and a novel marketing communication tool* that can be combined with traditional marketing tools. Therefore, the present paper seeks to investigate the relationship between content marketing and the five, traditional marketing communication tools to generate discussion if content marketing is the sixth element of the revised marketing communication mix.

**Literature review**

*Content marketing definition, functions, and spending*

Content marketing is the creation and distribution of relevant, timely, and valid content (Wang et al., 2017). Its primary purpose is to create customer trust and value (Repoviener, 2017). Content marketing may have entertaining or educational functions (Duc Le 2016; Lindström and Jörnéus, 2016). Content marketing can be effectively used both in B2C and B2B markets (Iankova et al. 2019). According to Kotler et al. (2017), the content can serve brand-building or sales promotion purposes. According to Moutsos (2019), 55% of the companies were capable of generating sales and income, and 53% of them were capable of increasing their existing customers' loyalty through content marketing in 2018. So, content marketing can be used to generate income and sales, and also, to increase customers' loyalty.

*Content types and formats*

Content marketing may appear in various formats based on the type of content. It could be audio and/or visual content (videos, live streaming, webinars); written digital content (articles, blogs, ebooks), images (infographics, photos, GIFs, charts), in-person content (events, presentations, workshops); audio-only digital content (podcasts, audiobooks), and written print content (magazines, books, brochures). Figure 1. shows the different types of content and how B2B marketers changed their use of content types/formats. Figure 2. shows the very same trends in B2C markets.

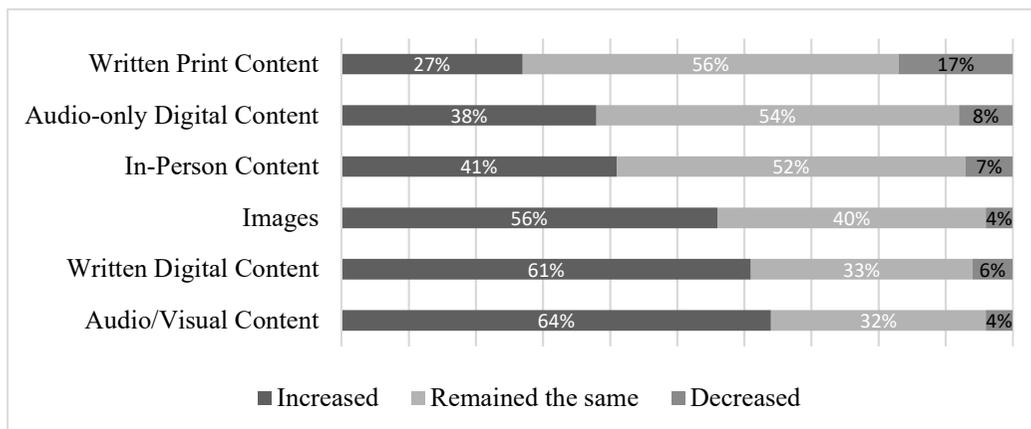

*Figure 1. The change of use of content types/format in B2B markets*
*Source: Own compilation based on Murton Beets, 2018*

As Figure 1 illustrates, in B2B markets, the use of audio/visual content; written digital content and images became more popular, while the use of written print content significantly decreased compared to the other types. The same trends can be seen in the B2C markets (Figure 2). The only slight difference between the two markets is in the use of audio-only digital content, which significantly dropped in the B2C market.

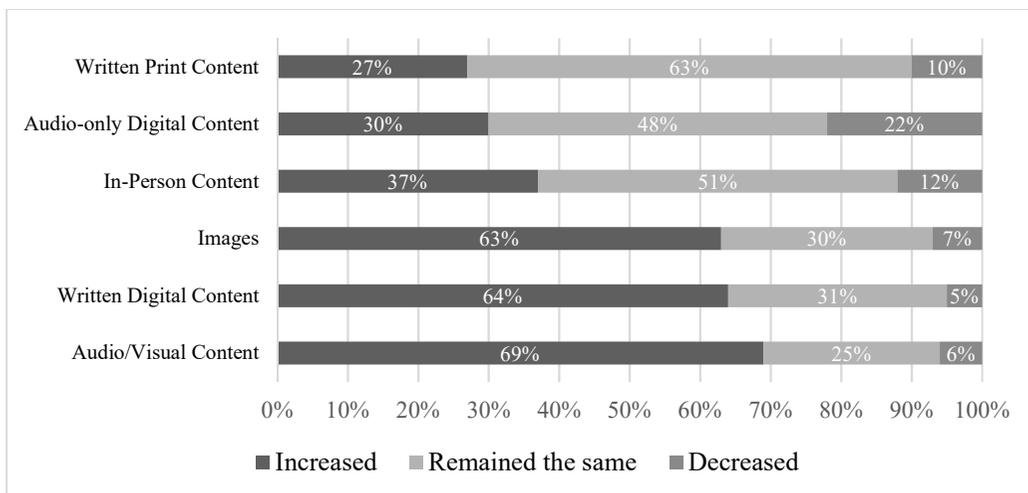

*Figure 2. The change of use of content types/format in B2C markets*
*Source: Own compilation based on Murton Beets, 2018*

In practice, various types of content can be used to reach out to consumers. As far as the type of content concerned, e-mail campaign is the most popular one, used by 87% of the companies (Murton Beets 2018). However, the following content types are also frequently used (values in brackets show the percentage of companies using the given content type): educative content (77%), actions calling for the next step (62%), events involving personal interactions (61%), telling stories (45%), offers (27%) and community building involving the public (23%). Trends and forecasts are less popular, only 5% of the companies used them (Murton Beets 2018).

*The goals of content marketing*

Content marketing helps to achieve several goals. The goal of content marketing is to gain customers (Barker, 2017) and to build customer relationships (Pažėraitė and Repovienė, 2018). Content marketing can very effectively be used to create brand awareness, educate audiences, generate demand/leads, and build credibility/trust (Figure 3.).
Also, content marketing is an effective tool for nurturing subscribers/audience/leads; driving attendance to one or more in-person events, building loyalty with existing clients, and supporting the launch of a new product. It can even be used to achieve sales/revenue generation and build a subscribed audience. Figure 3. presents the possible goals companies managed to successfully achieve by using content marketing.

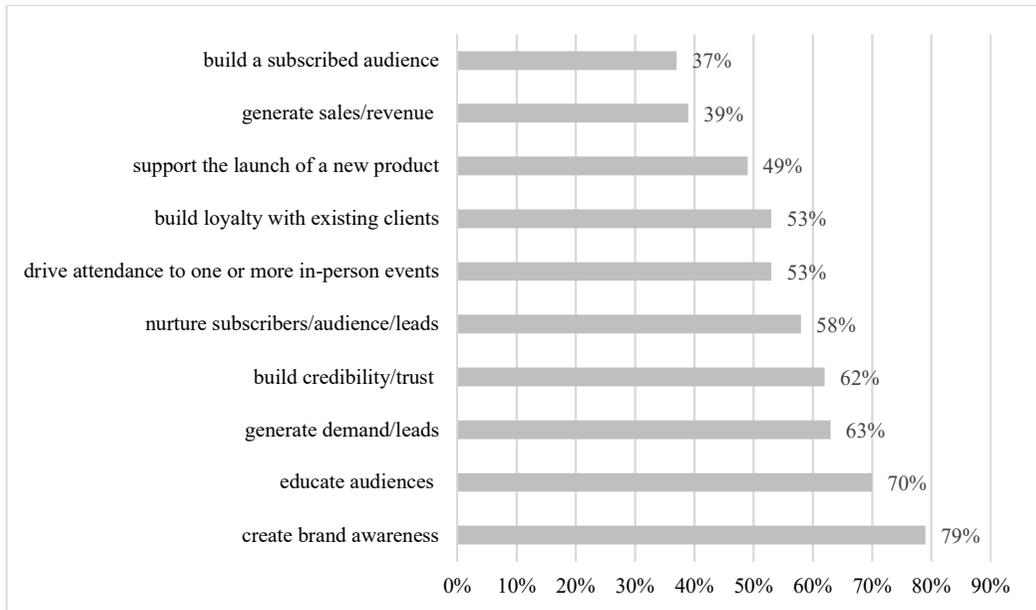

*Figure 3. Content marketing goals*
*Source: Own compilation based on Content Marketing Institute (2019).*
Note: Goals enterprise marketers have achieved by using content marketing successfully

**Research methodology**

This paper seeks to generate a debate on the current state of content marketing, and it aims to create a base for future quantitative research. It synthesizes the relevant literature to analyze the relationship between content marketing and the traditional marketing communication tools. It makes an attempt to distinguish content marketing from the other elements in marketing communication mix, which are advertising, sales promotion (SP), public relations (PR), personal selling, and direct marketing (DM). In the following section, based on extensive literature review, the five traditional marketing communication tools are compared to content marketing to reveal the similarities and differences between them regarding the type, purpose, standardization, time span and reach of communication and the target groups.

**Research findings and discussion**

*The relationship between advertising and content marketing*

Advertising is the most prominent element of the traditional communication mix. According to Horváth and Bauer (2013) advertising is an impersonal form of communication that reaches out to the recipients through mass media. Advertising mainly focuses on the product, specific product features, added services, price, packaging unit, trademark, logo, value and ideas worth considering from a social point of view (CSR). Kotler and Keller (2012) are committed to a narrower interpretation of advertising stating that advertising is only related to products, brands and/or services. In advertising, recipients (target group members) are usually aware of the fact that the main intention of marketers with the ads is to persuade and influence their behaviour. Since companies use advertising channels to relay commercials, their target group members can be reached indirectly. In this respect, content marketing is quite different. According to Kotler et al (2017), content marketing communicates with the marketer's own public. Content marketing also has an appropriately distinguished and defined target audience that receives more personalized content (Hajdú 2018).

Kotler et al (2017) express that the concept of traditional media is "one to many", while content marketing, especially social media, almost always mean two-way interactions. Furthermore, advertising helps to sell the product, while content marketing helps the customers to solve their problems and achieve their individual goals. According to Kotler et al (2017), consumers are ready to share the content, while the traditional ads, which are limited in time and space, are rather "skimmed over" by the target audience. It is almost sure to say that advertisements disturb a lot of people since they interrupt their favorite series, or delay videos they want to watch instantly, or fill their mailboxes with emails. Therefore, we can conclude that advertising has an intervening feature. Content marketing aims to maintain a lasting relationship with the target population (Pažėraitė and Repovienė 2018), while advertising is often seasonal and campaign-based (Kotler-Keller, 2012). Table 1 illustrates the main differences between advertising and content marketing. So, as Scott (2013) concluded, marketers can buy attention (advertising) or can own attention by creating something interesting and valuable that is published online for free (content marketing).

*Table 1.: The comparison of the traditional advertising and content marketing*

|  | traditional advertising | content marketing |
|---|---|---|
| **type of communication** | one-way: "I speak only" | two-way: "let's talk" |
| **purpose of communication** | promotion of products, brands and services | solving the customer's problem at no cost |
| **perception of communication from the customer's viewpoint** | intervening, disturbing | giving a helping hand |
| **reach** | a wide range of the population | individuals or groups |
| **standardization level** | standardized and impersonal | specified and more personalized |
| **target groups** | not own | own |
| **time span of communication** | short and campaign-based | a lasting relationship |
| **limitation** | limited | free |
| **target group reaction** | rejection, skimming over | sharing |

*Source: Own compilation based on Kotler et al, 2017; Horváth and Bauer, 2013; Hajdú, 2018; Maczuga et al, 2015; Pažėraitė and Repovienė, 2018*

*The relationship between direct marketing and content marketing*

Direct marketing (DM) is an addressed and interactive form of communication. It aims to achieve measurable responses, which can be orders, purchases, inquiries, or donations. Direct marketing is essentially built on databases. "It allows the potential customers to obtain information, it helps to establish the popularity of a brand or induces immediate purchases" (Horváth and Bauer, 2013, pp. 242.). The fact that direct marketing is built on databases implies that the customer value can be targeted quite accurately. Also, this marketing communication tool is easily optimizable. Telemarketing, mail advertisement, direct mail and direct response advertising are the forms of direct marketing (Horváth and Bauer, 2013).

Building brand awareness and credibility are definitely a common point in direct marketing and content marketing. However, direct marketing is less digital than content marketing. In general, the internet as a medium is less dominant in direct marketing, except for e-mail marketing. The purpose of communication in direct marketing is to present the product to make bids. Therefore, direct marketing is usually related to selling (receiving orders); the eye-catching presentation of products (catalogs) and advertising (mail advertisement).
According to Tapp (1999, pp. 23) "direct marketing is rather a sales system than a communication tool". Although nowadays direct marketing has widely been accepted as a marketing communication tool, its sales function cannot be ignored. This point of view is also appeared in Kotler and Keller (2012). According to Horváth and Bauer (2013), direct marketing provides the recipient with a clear opportunity to respond and directly targets the previously defined target groups. Although it also has a pre-defined target group (Hajdú 2018), content marketing places less emphasis on the sales-related responses. In content marketing, the responses affect the content itself. In content marketing, building trust, solving the customer's problem and providing further contents contribute to initiating purchases (Barker 2017).
There is another significant difference between direct marketing and content marketing. Direct marketing advertises a product or a service in a targeted manner to increase sales volume through immediate selling. That is why direct marketing is also called "direct order marketing", or "direct advertising". Consequently, direct marketing focuses only on the product, which offers the value for the customer (Kotler and Keller, 2012). Content marketing creates value and provides consumers with it. However, content marketing does not aim to sell immediately, only in one step (Fivetechnology, 2019), it has got longer time-orientation.
Combining direct marketing with content marketing can be very effective. If a customer registers an account online, he or she can receive free content (e.g. an ebook), which is content marketing, however, the data provided during the registration are also used to build a database, which can be used for direct marketing purposes. Content marketing that builds an audience not only identify demands but also generate it.

*The relationship between personal selling and content marketing*

Few researchers have addressed the question how personal selling and content marketing can be connected. Personal selling is a face-to-face selling technique where the emphasis is on personal interaction. In an event, which can be related to personal selling or could be a content marketing format, the company (brand) and its potential and existing customers can meet in person and/or online. However, it is important to note that the event is only one of several content marketing types, which are mostly digital.
Nowadays, the theory of selling as the most important task of the sales staff has already become outdated since the sales department is usually responsible for many other tasks, such as searching for potential customers, providing information, choosing the target market, providing services, collecting information and distribution (Kotler-Keller 2012, pp. 637).
Information that the sales staff provide about the products and services, in principle, can refer to the content marketing. Furthermore, services can also link personal selling and content marketing when the sales personnel try to solve the customer's problem.
Personal selling and content marketing can sometimes be combined but they can hardly be fell into one category due to the fundamental differences in their characteristics.

*The relationship between public relations and content marketing*

Content marketing should not be confused with public relations (Percy, 2018). In many cases, content marketing is a communication form used on a regular (daily or weekly) bases (Insights 2018). Content marketing aims to be part of the consumer's life and seeks to provide value to the customers in an educating and entertaining manner (Lindström and Jörnéus, 2016).
Public relation (PR) is a strategic tool aiming to turn brand messages into stories that are appealing to the media and its target audiences (Konczosné Szombathelyi 2018). Thus, PR builds credibility and trust among the stakeholders (Horváth-Bauer 2013). Since public relations is not sales-oriented, it is the changes in the mindset of the target audience that should be measured, not its effects on sales (Józsa et al, 2005). PR seeks to build a good reputation of the company; promote the success of the brand; deals with counselling and consulting. All these goals are very similar to those of content marketing, which among other things aims to build credibility and trust. However, content marketing is not a replacement for public relations (Mathewson and Moran, 2016)

Józsa et al (2005) emphasize that whatever the goal of PR is, the focus should be on creating trust by emphasizing understanding and willingness to cooperate to gain support from the stakeholders of the company. Trust is also a key factor in building strong brands. Both PR and content marketing can be regarded as regular and systematic communication activities (Józsa et al 2005; Muotsos 2017), and both use rather similar tools such as articles, newsletters, blogs, publications, social media, statistics, e-books, events, etc. (Probusiness, 2018).

However, there are some differences between PR and content marketing. Although trust is essential in PR, counselling is only a PR tool or technique. Counseling in PR refers to how we communicate with our clients. It is a recommended course of action that will serve the client's goals. On the contrary, in content marketing, the valuable content is always provided in the form of education, relevant information or entertainment (Lindström és Jörnéus, 2016).

The problem of measuring the effect of PR on sales is also a major difference. The impact of content marketing on sales is a lot easier to measure (Hajdú, 2018), moreover, one of the explicit goals of content marketing is to convert the target public into customers (Barker 2017). Effectiveness of content marketing can easily be measured due to its digital nature. Content marketing is customer-centred, focuses only on selected stakeholders and seeks to solve the customer's problem by providing information or educational content in an entertaining way (Lindström and Jörnéus 2016; Duc Le 2016). In content marketing, the goal is not to provide all the information but only the relevant content (Wang et al 2017). According to Hajdú (2018), content marketing is a profit-oriented tactical activity to gain customers and make deals. This means that content marketing acquires customers within a reasonable time-period. Content marketing not only produces content, but it also distributes it through its own channels, whereas PR works quite differently in this respect.

It is advisable is to combine content marketing with PR since they complement each other. PR can help marketers to make a better story about the brand (Spencer 2014).

*The relationship between sales promotion and content marketing*

There is a scarcity of literature devoted to the analysing the relationship between sales promotion and content marketing. Horváth and Bauer (2013) refer to sales promotion as a direct influence on consumer behaviour and an impetus to action. With reference to Bauer and Berács (2006), they emphasize that the primary goal of sales promotion is to promote product sales. "Sales promotion is a set of short-term incentive tools which aim to make consumers purchase more products more frequently or buy specific products or services" (Kotler-Keller, 2012, pp. 596). Regarding the consumer's benefit, sales promotion tools can be divided into two categories. Utilitarian and hedonistic tools can be distinguished. The utilitarian tools provide financial benefits (e.g. price discounts), whereas the hedonistic tools are focusing on entertainment, customer experience and loyalty (Yeshin, 2006). Product samples, gifts, contests and events (trade shows and exhibitions), the tools of sales promotion used to create the customer experience (hedonism), are very much related to content marketing (Józsa, 2014).

Product samples make it easier for the customers to try the products. It is an important link between content marketing and sales promotion because content marketing also provides customers with free and useful content when offering a solution to the customer's problem. Thereby, the company can demonstrate its competence and excellence by offering the best solutions to the customer's problem. Gifts are also commonly used in content marketing in the form of free content. On the contrary, gifts in sales promotion are not free, they are only given to the customers after the purchase (Horváth and Bauer, 2013).

Events can belong to content marketing and sales promotion; and sometimes even to PR, depending on their goals and their implementation (Józsa et al 2005; Dankó, 2008, Kranz-Pulizzi, 2011). The content of the event is the decisive factor. In an event, if marketers present information about how the customer could solve his or her problem, it is highly likely to be content marketing. Contests, games or phone applications can also be content marketing tools (Kranz-Pulizzi, 2011). However, contests and games as sales promotion tools are commonly used to increase sales. In this latter case, purchase is often a pre-requisite of entering the contest.

In sales promotion, the customer experience is directly linked to the purchase, while in content marketing it is not the case. According to Yeshin (2006), in sales promotion, customer loyalty is gained through financial benefits and consumption (e. g. through loyalty points, gifts, etc.). On the contrary, content marketing seeks to achieve the same goal by providing free content that is useful and/or entertaining. The primary goal of sales promotion is to increase product sales, which can be a distinguishing factor between content marketing and sales promotion. Although content marketing is also sales-oriented in the long run, here the deal is achieved in several steps (Fivetechnology, 2019). In this process, the very first step is building trust by giving value without asking for compensation or purchase (Repoviener, 2017; Maczuga et al, 2015). We can conclude that the main differences between content marketing and sales promotion can be found in their objectives and time-orientation. Content marketing, which is not a short-term tool, is often regarded as is an introductory stage of sales as it does not aim to make purchases quickly.

**Conclusions**

This paper investigates the relationship between content marketing and the five traditional marketing communication tools. The goal of the article is to generate a discussion on the status of content marketing. In this paper, content marketing was compared to advertising, direct marketing, personal selling, public relations and sales promotion to find out the main differences and similarities. An extensive literature review explored some fundamental differences between the traditional marketing communication tools and content marketing. Based on this result, *content marketing can be regarded as a novel marketing communication tool and the sixth element of the revised marketing communication mix*. Content marketing can be effectively used in marketing campaigns in the digital environment. Because of its digital nature, content marketing can be more effective in digitally advanced target markets. One of the positive effects of COVID-19 is the accelerated digitalization, which favorable to the use of content marketing.

**Acknowledgements**

"The described article/presentation/study was carried out as part of the EFOP-3.6.1-16-2016-00011 "Younger and Renewing University – Innovative Knowledge City – institutional development of the University of Miskolc aiming at intelligent specialisation" project implemented in the framework of the Szechenyi 2020 program. The realization of this project is supported by the European Union, co-financed by the European Social Fund."

"A cikkben/előadásban/tanulmányban ismertetett kutató munka az EFOP-3.6.1-16-2016-00011 jelű „Fiatalodó és Megújuló Egyetem – Innovatív Tudásváros – a Miskolci Egyetem intelligens szakosodást szolgáló intézményi fejlesztése" projekt részeként – a Széchenyi 2020 keretében – az Európai Unió támogatásával, az Európai Szociális Alap társfinanszírozásával valósul meg"

**References**

BARKER, S. (2017). *How to Create High-Converting Content*. Retrieved from https://contentmarketinginstitute.com/2017/05/create-high-converting-content, 10.04.2019.

BRENNER, M. (2019). *Content Marketing Survey: Marketers Focus On Content Creation.* Retrieved from: https://marketinginsidergroup.com/content-marketing/2019-content-marketing-survey-content-creation/, 11.04.2019.

CONTENT MARKETING INSTITUTE (2019). *Enterprise Content Marketing 2019 – Benchmarks, Budgets, and Trends – North America*. Retrieved from https://contentmarketinginstitute.com/wp-content/uploads/2019/02/FINAL-2019_Enterprise_Research.pdf, 05.04.2019.

DANKÓ, L. (2008). *Értékesítés-ösztönzés.* Marketing Intézet Miskolc. Miskolci Egyetem. Pro Marketing Miskolc Egyesület

DUC LE, M. (2013). *Content Marketing*, Haaga-Heila University of Applied Sciences, Porvoo

FIVETECHNOLOGY (2019). *Content Marketing & Strategy*. Retrieved from: https://www.fivetechnology.com/internet-marketing/content-marketing-strategy, 02.03.2019.

HAJDÚ, N. (2018). Az online marketingcontrolling értékelési folyamata a tartalommarketing ROI segítségével. *Controller Info* 6 : 1 pp. 5-8. , 4 p. 10.24387/CI.2018.1.2

HORVÁTH, D. - BAUER, A. (2013). *Marketingkommunikáció – Stratégia, új média, fogyasztói részvétel.* Budapest: Akadémiai Kiadó.

IANKOVA S. - DAVIES I. - ARCHER-BROWN C. - MARDER B. - YAU A. (2019). A comparison of social media marketing between B2B, B2C and mixed business models. *Industrial marketing management.*, 81, 169-179. doi:10.1016/j.indmarman.2018.01.001

JÓZSA, L., PISKÓTI, I., REKETTYE, G., VERES, A. (2005). Döntésorientált marketing. KJK KERSZÖV Jogi és Üzleti kiadó Kft. Budapest

KAMARAONLINE (2018). *Versenyhátrányt okoz a magyar vállalkozásoknak az informatikai lemaradás*. Retrieved from: http://kamaraonline.hu/cikk/versenyhatranyt-okoz-a-magyar-vallalkozasoknak-az-informatikai-lemaradas, 30.04.2018.

KONCZOSNÉ SZOMBATHELYI, M. (2018). *A PR úttörői és napjainkig tartó hatásuk*, Széchenyi István Egyetem, Győr, Retrieved from: https://kgk.sze.hu/images/dokumentumok/VEABtanulmanyok/konczosne_szombathelyi_marta.pdf, 05.04.2019.

KOTLER, P. - KELLER, K., L. (2012). *Marketingmenedzsment*. Budapest: Akadémiai Kiadó.

KOTLER, P. - KARTAJAYA H. - SETIAWAN I. (2017). *Marketing 4.0 - Moving from Traditional to Digital.* John, & Sons, Inc.

KRANZ J., PULIZZI J. (2011). *Content Marketing Playbook*, Junta42 Conent Marketing Institute & Kranz Communications

LINDSTRÖM, A. L. - JÖRNÉUS, A. (2016). *Co-Creating value through Content Marketing*, University of Gothengurg, School of Business, Economics and Law

MACZUGA P. - SIKORSKA K. (2015). *Content marketing Handbook: Simple Ways to Innovate Your Marketing Approach*. Project under Lifelong Learning Programme of European Commission. Warsaw. Racom Communications. ISBN: 978-83-63481-10-0

MATHEWSON, J. - MORAN, M. (2016). *Outside-in marketing: Using big data to guide your content marketing.* Boston: IBM Press, Pearson plc

MOUTSOS, K. (2017). *Publishing frequency: why (and how) we're changing things up.* Retrieved from: https://contentmarketinginstitute.com/2017/12/publishing-frequency-changing/ 10.04.2019.

MOUTSOS, K. (2019). *Tech Content Marketers Talk Content Creation Challenges, Tools, and Trends* Retrieved from: https://contentmarketinginstitute.com/2019/03/tech-content-marketers-research/, 10.04.2019.

MURTON BEETS, L. (2018). *2019 B2B Content Marketing Research: It Pays to Put Audience First.* Retrieved from: https://contentmarketinginstitute.com/2018/10/research-b2b-audience/, 11.04.2019.

NAGY, L. (2020). The Neuromarketing Analysis and the Categorization of Television Commercials, *Észak-magyarországi Stratégiai Füzetek,* XVII. évf., 2020, 2., pp. 79-88. https://doi.org/10.32976/stratfuz.2020.16

NAGY, S. (2017). The Impact Of Country Of Origin In Mobile Phone Choice Of Generation Y And Z., *Journal Of Management And Training*, Vol.4. No. 2., pp. 16-29. https://doi.org/10.12792/JMTI.4.2.16

NAGY, S. (2016). E-commerce in Hungary: A Market Analysis, *Theory Methodology Practice*: *'Club of Economics in Miskolc'*, Vol.12., Nr.2, pp. 25-32. http://dx.doi.org/10.18096/TMP.2016.03.03

PISKÓTI, I. (2018). *Digitalizáció – az új marketingkoncepció és stratégiai megoldások irányai – Marketing 4.0, A digitalizáció és annak társadalmi-gazdasági hatásai*, Gazdálkodástudományi Bizottság konferenciája, Marketingtudományi Szekció, Budapesti Corvinus Egyetem, 2018. nov. 13. Retrieved from: https://docplayer.hu/108312523-Digitalizacio-az-uj-marketingkoncepcio-es-strategiai-megoldasok-iranyai-marketing-4-0.html, 17.07.2020

PAŽĖRAITĖ, A. - REPOVIENĖ, R. (2018). Content Marketing Decisions for Effective Internal Communication. *Management of Organizations: Systematic Research*, Sciendo, vol. 79(1), pp. 117-130., doi:10.1515/mosr-2018-0008

PERCY, L. (2018). *Strategic integrated marketing communications.* London: Routledge.

PROBUSINESS (2018). *A tartalommarketing is PR?* Retrieved from: https://www.probusiness.hu/pr/a-tartalommarketing-is-pr/, 09.04.2019.

REPOVIENER, R. (2017). Role of content marketing in a value creation for customer context: a theoretical analysis, *International Journal of Global Business Management and Research*, Vol 6., issue 2., p. 37-48

SCOTT, D. M. (2013). *The new rules of marketing & PR: How to use social media, online video, mobile applications, blogs, news releases, & viral marketing to reach buyers directly*. Hoboken, NJ: John Wiley & Sons.

SETHI, M. (2018). *Social Media Strategy for Content Marketing*, Retrieved from: https://www.entrepreneur.com/article/321683, 02.12.2018.

SPENCER, J. (2014). *The Role of PR and Content Marketing in 2015.* Retrieved from: https://contentmarketinginstitute.com/2014/12/role-pr-content-marketing/, 10.04.2019.

YESHIN, T. (2006). *Sales Promotion*. London: Thomson Learning.

WANG, W. - MALTHOUSE, E. C. - CALDER, B. - UZUNOGLU, E. (2019). B2B content marketing for professional services: In-person versus digital contacts. *Industrial Marketing Management*, 81, 160-168. doi:10.1016/j.indmarman.2017.11.006